\documentclass[preprint,showpacs,showkeys,preprintnumbers,amsmath,amssymb]{revtex4}
\usepackage{amssymb}
\usepackage[dvips]{graphicx}

\begin{document}

\title{Residual Symmetry Reductions and Interaction Solutions of (2+1)-Dimensional Burgers Equation}

\author{ Xi-zhong Liu, Jun Yu, Bo Ren }

\affiliation{Institute of Nonlinear Science, Shaoxing University, Shaoxing 312000, China}
\begin{abstract}
The (2+1)-dimensional Burgers equation has been investigated first from prospective of symmetry by localizing the nonlocal residual symmetries and then studied by a simple generalized tanh expansion method. New symmetry reduction solutions has been obtained by using the standard Lie point symmetry group approach. A new B\"{a}klund transformation for Burgers equation has been given with the generalized tanh expansion method . From this BT, interactive solutions among different nonlinear excitations which is hard to obtain by other methods has also been obtained easily.
\end{abstract}



\maketitle
\section{Introduction}
Due to the important role that nonlinear theory plays in mathematics, physics, chemistry, biology, communications, astrophysics and geophysics, much endeavor have been devoted to find important and significant properties of different type of nonlinear equations. Among which, the study on integrable systems occupies the most important place. The integrable systems are introduced by scientist with some ideal conditions to describe complex nonlinear phenomenon, such as the Korteweg-de Vries (KdV) equation \cite{kdv}, the nonlinear Schr\"{o}dinger equation (NLS) \cite{sulem}, the sine-Gordon (SG) equation \cite{rubin}, Davey-Stewartson (DS) equation \cite{davey}, the Kadomtzev-Pedviashvili (KP) [5] equation and so on. Various effective methods like the inverse scattering transformation (IST), Darboux transformation, B\"{a}cklund transformation (BT)\cite{matv}, Hirota's direct method \cite{hiro}, mapping and deformation approach \cite{lou1}, standard and extended truncated Painlev\'{e} analysis \cite{lou2} etc. have been well developed to find exact solutions for integrable models.

Since the Lie group theory was introduced by Sophus Lie to study differential equations \cite{lie}, the symmetry theory has been developed a lot for studying nonlinear equations. It is well known that using Lie point symmetries, the dimensions of nonlinear equations can be reduced through classical or non-classical Lie group approaches and the exact solutions of these equations can be constructed further. However, for integrable system there exist infinitely many non-local symmetries which are linked with potential symmetries \cite{blun}, inverse recursion operators \cite{lou3}, negative hierarchies, conformal invariance, Darboux transformations, B\"{a}klund transformations, and so on. It would be very interesting to use these non-local symmetries for studying important properties of the corresponding equations. An direct thinking is to localize these symmetries by prolong the original system then use the traditional Lie point symmetry theory to reduce the prolonged system \cite{huxiao} and find new exact reduction solutions which is usually hard to be found by traditional DT or BT method.

The Painlev?e analysis is usually used to study the integrability of nonlinear systems. Painlev?e integrable demands that there exist Laurent series on open sets of the complex time variable solutions and these solutions are consistent, or match on the overlapping pieces of the sets on which they are defined. In other words this notion of integrability is existence of meromorphic solution. When the equations has Painlev\'{e} property, the truncated Painleve expansion method can be used to find exact solutions of the equations. The recent studies on Painleve analysis reveals that the truncated Painlev\'{e} expansion is the backlund transformation and the residue of the expansion is just the non-local symmetry related this B\"{a}cklund transformation, which is called residual symmetry. In Ref. \cite{chengxue}, the truncated Painlev\'{e} expansion method is further developed to the general the hyperbolic tangent (tanh) function expansion approach by allowing the expansion coefficients being arbitrary functions of space-time and interaction solutions between solitons and any Schr\"{o}dinger waves has been obtained. Unlike the usual tanh expansion method which lose many essential information of the original equation the generalized tanh expansion method can be used to find much more general solutions retrieving the missing essential properties.

In this
paper, we concentrate on the (2+1)-dimensional Burgers equation with constants $a,b$ in the form
\begin{subequations}\label{bur}
\begin{equation}\label{bur1}
u_t=uu_y+avu_x+bu_{yy}+abu_{xx},
\end{equation}
\begin{equation}\label{bur2}
u_x=v_y
\end{equation}
\end{subequations}
via residual related symmetry reduction and generalized tanh function expansion approach to obtain new BT transformations and interaction solutions between different solitons. An equivalent form of the Burgers Eq. (2) has been derived from the generalized Painlev\'{e} integrability classification
in Ref. \cite{hong}.

Burgers equation is a fundamental partial differential equation in the field of fluid mechanics. It also occurs in various areas of applied physics such as the modeling of gas dynamics and traffic flow. The exact solutions of Burgers equation have been studied by many authors. In Ref. \cite{lou4}, the infinite many symmetries and exact solutions of Burgers equation are obtained by the repeated
symmetry reduction approach.

In Sec. 2 of this paper, the residual symmetry of Burgers system is localized and then Levi transformation (the second DT) is re-obtained via Lie's first theorem. In Sec. 3, the localized symmetry group is analyzed and classified according to different transformation properties. The reduction solutions related with space-time and Mobious transformation invariance symmetries has also been obtained subsequently. In Sec. 4, the generalized tanh expansion approach is used to Burgers equation to obtain new B\"{a}cklund transformation. Using the BT, some special explicit novel exact solutions and the interactions between them has also been investigated. It is found that completely non-elastic interactions and completely elastic interactions will occur when specific parameters be fixed. The last section is devoted to a short summary and discussion.

\section{Localization of the non-local residual symmetries of the Burgers equation}

By means of the Painlev¡äe analysis, various integrable properties such
as the B¡§acklund transformations, Lax pairs, infinitely many symmetries, bilinear forms, and so on, can be easily found if the
studied model possesses Painlev¡äe property, i.e., it is Painlev¡äe integrable.

For the Burgers Eq. \eqref{bur}, the truncated Painlev\'{e} expansion takes the general form
\begin{equation}\label{genpain}
u=\sum_{i=0}^{\alpha}{u_if^{i-\alpha}},~ v=\sum_{j=0}^{\beta}{v_jf^{j-\beta}},
\end{equation}
where $u_\alpha$ and $v_\beta$ are arbitrary solutions of the discussed system while $u_{\alpha-1},u_{\alpha-2},\cdots, u_0$ and $v_{\beta-1},v_{\beta-2},\cdots,v_0$ are all related to derivatives of $f$. In order to determine $\alpha$ and $\beta$, we first substitute $u\sim u_0f^\alpha$ and $v\sim v_0f^\beta$ into Eq. \eqref{bur} and balance the nonlinear and dispersion terms to give $\alpha=1,\ \beta=1$. Hence the truncated Painlev\'{e} expansion reads
\begin{equation}\label{pain1}
u=\frac{u_0}{f}+u_1,\, v=\frac{v_0}{f}+v_1.
\end{equation}
It can be easily proved that the residuals $u_0,v_0$ are the
symmetries corresponding to the BT[].

With the property of Painlev\'{e} integrable the BT \eqref{pain1} transforms the original Burgers system \eqref{bur} into its Schwarzian form
\begin{equation}\label{sch}
S(f)=0.
\end{equation}
That mens \eqref{sch} is form invariant under the Mobious transformation
\begin{equation}
f\to \frac{a_1f+b_1}{a_2f+b_2},\, a_1a_2 \neq b_1b_2
\end{equation}
which means \eqref{sch} possess three symmetries $\sigma^f=d_1$,$\sigma^f=d_2f$ and
$$\sigma^f=d_3f^2$$
with arbitrary constants $d_1,\,d_2$ and $d_3.$

Substituting Eq. \eqref{pain1} into Eq. \eqref{bur} yields
\begin{multline}\label{subspain1}
-abu_{1,xx}-av_1u_{1,x}-u_1u_{1,y}+u_{1,t}-bu_{1,yy}+f^{-1}(-av_1u_{0,x}-av_0u_{1,x}
-abu_{0,xx}\\-u_1u_{0,y}+u_{0,t}-u_0u_{1,y}-bu_{0,yy})+f^{-2}(av_1u_0f_x+2abu_{0,x}f_x
+abu_0f_{xx}+bu_0f_{yy}-av_0u_{0,x}\\+2bu_{0,y}f_y-u_0u_{0,y}+u_1u_0f_y-u_0f_t)-f^{-3}u_0
(2abf_x^2-av_0f_x+2bf_y^2-u_0f_y)=0,
\end{multline}
and
\begin{equation}\label{subspain2}
u_{1,x}-v_{1,y}+f^{-1}(-v_{0,y}+u_{0,x})+f^{-2}(v_0f_y-u_0f_x)=0
\end{equation}
Vanishing the coefficients $f^{-3}$ yields
\begin{equation}\label{u0v0}
u_0 = 2bf_y,\, v_0 = 2bf_x
\end{equation}
Substituting Eq. \eqref{u0v0} into Eq. \eqref{subspain2}
yields the Burgers equation \eqref{bur2} with $u_1$ and $v_1$ as solutions
\begin{equation}\label{u1x}
u_{1,x}-v_{1,y}=0.
\end{equation}
Substituting Eq. \eqref{u0v0} into Eq. \eqref{subspain1} and vanishing the coefficient of $f^{-2}$
yields
\begin{equation}\label{ft}
av_1f_x+abf_{xx}+bf_{yy}+u_1f_y-f_t=0,
\end{equation}
while the coefficient of $f^{-1}$ equals Eq. \eqref{u1x} by using Eq. \eqref{ft}.

Vanishing the coefficient of $f^0$ in Eq. \eqref{subspain1}, we have
\begin{equation}\label{u1bur1}
u_{1,t}-abu_{1,xx}-av_1u_{1,x}-u_1u_{1,y}-bu_{1,yy}=0
\end{equation}
which is just the Burgers equation \eqref{bur1} with the solution $u_1$ and $v_1.$

Now, using the standard truncated Painleve expansion
\begin{eqnarray}\label{bt1tran}
\nonumber u=\frac{2bf_y}{f}+u_1,\\
v=\frac{2bf_x}{f}+v_1,
\end{eqnarray}
we have the following B\"{a}cklund transformation theorem

\noindent\textbf{\emph{Theorem 1}.} \eqref{bt1tran} is a BT between the solutions $\{u,\, v\}$ and $\{u_1,\, v_1\}$ if the latter is related to $f$ by Eq. \eqref{ft}.

Since $u_0$ and $v_0$ are non-local symmetries corresponding to BT, one can naturally believe that it can be localized to a Lie point symmetry such that we can use the Lie' s first theorem to recover the original BT. To this end, we introduce new variables to eliminate the space derivatives of $f$
\begin{equation}\label{gh1}
g\equiv f_x,
\end{equation}
\begin{equation}\label{gh2}
 h\equiv f_y
\end{equation}
and investigate the symmetries relations between different variables. That is to say, we have to solve the linearized system of the prolonged equations of \eqref{bur}, \eqref{ft},\eqref{gh1} and \eqref{gh2}
\begin{subequations}\label{linear}
\begin{equation}
\sigma_{u_1,x}-\sigma_{v_1,y}=0,
\end{equation}
\begin{equation}
-ab\sigma_{v_1,xy}+\sigma_{u_1,t}-b\sigma_{u_1,yy}-\sigma_{u_1}u_{1,y}-u_1\sigma_{u_1,y}-a\sigma_{v_1}v_{1,y}
-av_1
\sigma_{v_1,y}=0,
 \end{equation}
 \begin{equation}
 \sigma_{f,x}-\sigma_g=0,
\end{equation}
\begin{equation}
  \sigma_{f,y}-\sigma_h=0,
\end{equation}
\begin{equation}
 \sigma_{f,t}-b\sigma_{f,yy}-av_1\sigma_{f,x}-a\sigma_{v_1}f_x
-ab\sigma_{f,xx}-f_y\sigma_{u_1}-u_1\sigma_{f,y}=0
\end{equation}
\end{subequations}
It can be easily verified that the solution of \eqref{linear} has the form
\begin{subequations}\label{pointsy1}
\begin{equation}
\sigma^{u_1} = 2bh,
\end{equation}
\begin{equation}
\sigma^{v_1} = 2bg,
\end{equation}
\begin{equation}
\sigma^g = -\frac{fg}{b},
\end{equation}
\begin{equation}
\sigma^h = -\frac{fh}{b},
\end{equation}
\begin{equation}
\sigma^{f} = -\frac{f^2}{2b},
\end{equation}
\end{subequations}
if $d_3$ is fixed as $-\frac{1}{2b}$.

The result \eqref{pointsy1} indicates that the residual symmetries \eqref{u0v0} is localized in
the properly prolonged system \eqref{bur}, \eqref{ft}, \eqref{gh1} and \eqref{gh2} with the Lie point symmetry vector
\begin{equation}\label{pointV}
V=2bh\partial_{u_1}+2bg\partial_{v_1}-\frac{fg}{b}\partial_{g}-\frac{fh}{b}\partial_h
-\frac{f^2}{2b}\partial_f.
\end{equation}
In other words, symmetries related to the truncated Painlev\'{e} expansion is just a special Lie point symmetry of the prolonged system  \eqref{bur}, \eqref{ft}, \eqref{gh1} and \eqref{gh2}.

Now has we obtained the localized residual symmetries, the interesting question is to ask what kind of finite transformation would be corresponding the Lie point symmetry \eqref{pointV}. Then we has the following theorem.

\noindent\emph{ \textbf{Theorem 2.}}
If $\{u_1,v_1,g,h,f\}$ is a solution of the prolonged system \eqref{bur}, \eqref{ft}, \eqref{gh1} and \eqref{gh2}, then so is $\{\hat{u_1},\hat{v_1},\hat{g},\hat{h},\hat{f}\}$ with
\begin{subequations}\label{pointsy}
\begin{equation}
\hat{u_1}= u_1+\frac{4hb^2\epsilon}{\epsilon f+2b},
\end{equation}
\begin{equation}
\hat{v_1} = v_1+\frac{4gb^2\epsilon}{\epsilon f+2b}
\end{equation}
\begin{equation}
\hat{g} = \frac{4gb^2}{(\epsilon f+2b)^2},
\end{equation}
\begin{equation}
\hat{h} = \frac{4hb^2}{(\epsilon f+2b)^2},
\end{equation}
\begin{equation}
\hat{f} = \frac{2bf}{\epsilon f+2b},
\end{equation}
\end{subequations}
with arbitrary group parameter $\epsilon$.

\emph{Proof.} Using Lie's first theorem on vector \eqref{pointV} with the corresponding
 initial condition as follows
\begin{eqnarray}
\\ \frac{d\hat{u_1}(\epsilon)}{d\epsilon}&=& 2b\hat{h}(\epsilon),\,\quad \hat{u_1}(0)=u_1,\\
\frac{d\hat{v_1}(\epsilon)}{d\epsilon}&=& 2b\hat{g}(\epsilon),\,\quad \hat{v_1}(0)=v_1,\\
\frac{d\hat{g}(\epsilon)}{d\epsilon}&=&-\frac{\hat{f}(\epsilon)\hat{g}(\epsilon)}{b},\,\quad \hat{g}(0)=g,\\
\frac{d\hat{h}(\epsilon)}{d\epsilon}&=&-\frac{\hat{f}(\epsilon)\hat{h}(\epsilon)}{b},\,\quad \hat{h}(0)=h,\\
\frac{d\hat{f}(\epsilon)}{d\epsilon}&=&-\frac{\hat{f}^2(\epsilon)}{2b},\,\quad \hat{f}(0)=f
\end{eqnarray}
one can easily obtain the solutions of the above equations stated in Theorem 1, thus the theorem is
proved.

Next let us consider the Lie point symmetry of the prolonged system in the general form
 \begin{equation}\label{vectorv1}
V=X\frac{\partial}{\partial x}+Y\frac{\partial}{\partial y}+T\frac{\partial}{\partial
t}+U_1\frac{\partial}{\partial u_1}+V_1\frac{\partial}{\partial v_1}+G\frac{\partial}{\partial g}+
H\frac{\partial}{\partial h}+F\frac{\partial}{\partial f},
\end{equation}
which means that the prolonged system is invariant under the following transformation
\begin{equation}
\{x,y,t,u,v,g,h,f\} \rightarrow \{x+\epsilon X,y+\epsilon Y,t+\epsilon T,u+\epsilon U,v+\epsilon V,g+\epsilon G,h+\epsilon G,h+\epsilon H,f+\epsilon F\}
\end{equation}
with the infinitesimal parameter $\epsilon$.
Equivalently, the symmetry in the form \eqref{vectorv1} can be written as a function form as
\begin{subequations}\label{sigmasy}
\begin{equation}
\sigma_{u_1} = Xu_{1x}+Yu_{1y}+Tu_{1t}-U_1,
\end{equation}
\begin{equation}
\sigma_{v_1} = Xv_{1x}+Yv_{1y}+Tv_{1t}-V_1,
\end{equation}
\begin{equation}
\sigma_g = Xg_{x}+Yg_{y}+Tg_{t}-G,
\end{equation}
\begin{equation}
\sigma_h = Xh_{x}+Yh_{y}+Th_{t}-H,
\end{equation}
\begin{equation}
\sigma_f = Xf_{x}+Yf_{y}+Tf_{t}-F.
\end{equation}
\end{subequations}
Substituting Eq. \eqref{sigmasy} into Eq. \eqref{linear} and eliminating $u_{1,t}, v_{1,y}, g_t, h_x, h_t, f_x, f_y$ and $f_t$ in terms of the prolonged system we get more than 250 determining equations for the functions $X, Y, T, U_1, V_1, G, H$ and $F$. Calculated by computer algebra, we finally get the desired result
\begin{eqnarray}\label{sol}
\nonumber&&X= \frac{c_1xt}{2}+\frac{c_2x}{2}+x_0, Y = \frac{c_1ty}{2}+\frac{c_2y}{2}+c_4t+c_5,\\\nonumber&& T= \frac{c_1t^2}{2}+c_2t+c_3,U_1 = -\frac{c_1y}{2}-\frac{c_1u_1t}{2}+c_6h-\frac{c_2u_1}{2}-c_4,\\\nonumber&& V_1 = -\frac{2x_{0,t}+av_1c_1t+av_1c_2-2ac_6g+c_1x}{2a}, G= -\frac{g(bc_1t+bc_2-2bc_7+2c_6f)}{2b},\\&& H = -\frac{h(bc_1t+bc_2-2bc_7+2c_6f)}{2b},F = -\frac{c_6f^2-2bc_7f-2bc_8}{2b}.
\end{eqnarray}
with arbitrary constants $c_1, c_2, c_3, c_4, c_5, c_6, c_7, c_8$ and arbitrary function $x_0$ of $t$.
The obtained symmetries can be categorized by different transformation invariance as follows,

(i) the residual symmetries and Mobious transformation by setting $c_1=c_2=c_3=c_4=c_5=x_0=0$ and
$c_6=c_7=c_8=-1$
\begin{equation}
\sigma_{u_1} = f_y, \sigma_{v_1} = f_x, \sigma_{g} = -\frac{(f-b)f_x}{b}, \sigma_{h} = -\frac{(f-b)f_y}{b}, \sigma_{f} = -\frac{f^2-2bf-2b}{2b}
\end{equation}

(ii) time transformation invariant symmetries by setting $c_1=c_2=c_4=c_5=c_6=c_7=c_8=x_0=0$ and
$c_3=1$
\begin{equation}
\sigma_{u_1} = u_{1,t}, \sigma_{v_1} = v_{1,t}, \sigma_{g} = g_t, \sigma_{h} = h_t, \sigma_{f} = f_t
\end{equation}

(iii) y-space transformation invariant symmetries by setting $c_1=c_2=c_3=c_4=c_6=c_7=c_8=x_0=0$ and
$c_5=1$
\begin{equation}
\sigma_{u_1} = u_{1,y}, \sigma_{v_1} = v_{1,y}, \sigma_{g} = g_y, \sigma_{h} = h_y, \sigma_{f} = f_y
\end{equation}

(iv) scaling  transformation invariant related symmetries by setting $c_1=c_3=c_4=c_5=c_6=c_7=c_8=x_0=0$ and
$c_2=2$
\begin{equation}\nonumber
\sigma_{u_1} = xu_{1,x}+yu_{1,y}+2tu_{1,t}+u_1, \sigma_{v_1} = xv_{1,x}+yv_{1,y}+2tv_{1,t}+v_1,\sigma_{g} = xg_{x}+yg_{y}+2tg_{t}+g,
\end{equation}
\begin{equation}
\sigma_{h} = xh_{x}+yh_{y}+2th_{t}+h, \sigma_{f} = xf_{x}+yf_{y}+2tf_{t}
\end{equation}

(v) time relevant x-space transformation invariant symmetries by setting $c_1=c_2=c_3=c_4=c_5=c_6=c_7=c_8=0$
\begin{equation}
\sigma_{u_1} = x_0u_{1,x}, \sigma_{v_1} = \frac{ax_0v_{1,x}+x_{0,t}}{a}, \sigma_{g} = x_0g_x, \sigma_{h} = x_0h_x, \sigma_{f} = x_0f_x
\end{equation}

(vi) Galileo transformation invariant symmetries by setting $c_1=c_2=c_3=c_5=c_6=c_7=c_8=x_0=0$ and
$c_4=1$
\begin{equation}
\sigma_{u_1} = 1+tu_{1,y}, \sigma_{v_1} =tv_{1,y}, \sigma_{g} = tg_y, \sigma_{h} = th_y, \sigma_{f} = tf_y
\end{equation}

(vii) time-Mobious transformation invariant symmetries by setting $c_2=c_3=c_4=c_5=c_6=c_7=c_8=x_0=0$ and
$c_1=2$
\begin{equation}\nonumber
\sigma_{u_1} = xtu_{1,x}+ytu_{1,y}+t^2u_{1,t}+tu_1+y, \sigma_{v_1} =xtv_{1,x}+ytv_{1,y}+t^2v_{1,t}+tv_1+\frac{x}{a},\end{equation} \begin{equation}\sigma_{g} = xtg_x+ytg_y+t^2g_t+gt, \sigma_{h} = xth_x+yth_y+t^2h_t+ht, \sigma_{f} =  xtf_x+ytf_y+t^2f_t
\end{equation}

Now let us proceed to reduce the prolonged system by standard Lie point symmetry group approach.
Without loss of generality, we consider the symmetry reductions related to space-time invariant and residual symmetries by setting $c_1=c_2=c_4=c_7=0$ and $c_3=c_5=c_6=c_8=x_0=1$. Substituting Eq. \eqref{sol} into Eq. \eqref{sigmasy} with Eqs. \eqref{gh1} and \eqref{gh2}, we get
\begin{eqnarray}\label{sigmauvf}
\nonumber\sigma_{u_1}&=&u_{1,x}+u_{1,y}+u_{1,t}-f_y,\\
\nonumber\sigma_{v_1}&=&v_{1,x}+v_{1,y}+v_{1,t}-f_x,\\
\sigma_f&=&\frac{1}{2b}(2bf_x+2bf_y+2bf_t+f^2-2b)
\end{eqnarray}
The corresponding group invariant solutions which can be obtained by solving Eqs. \eqref{u1x}, \eqref{ft}, \eqref{u1bur1} and \eqref{sigmauvf} with $\sigma_{u_1}=\sigma_{v_1}=\sigma_{f}=0$ has the general form
\begin{eqnarray}\label{redusol}
\nonumber f&=&\sqrt{2b}\tanh(\frac{x+F}{\sqrt{2b}}),u_1=\sqrt{2b}F_Y\tanh(\frac{x+F}{\sqrt{2b}})+U_1,\\
v_1&=&-\sqrt{2b}(F_Y+F_T-1)\tanh(\frac{x+F}{\sqrt{2b}})+V_1
\end{eqnarray}
where the group invariant variables are $Y=-x+y$ and $T=-x+t$, while $F,\, U_1$ and $V_1$ are all group invariant functions of $Y$ and $T$.

Substituting Eq. \eqref{redusol} into Eq. \eqref{bt1tran} we obtain the reduction solution for Burgers system \eqref{bur}
\begin{eqnarray}\label{redusoluv}
\nonumber
u&=&\sqrt{2b}F_Y\tanh^{-1}(\frac{x+F}{\sqrt{2b}})+U_1,\\
v&=&-\sqrt{2b}(F_Y+F_T-1)\tanh^{-1}(\frac{x+F}{\sqrt{2b}})+V_1
\end{eqnarray}

In order to get the symmetry reduction equations for the group invariant functions $F,\, U_1$ and $V_1$, we first substitute Eq. \eqref{redusol} into Eq. \eqref{u1x} and find
\begin{equation}\label{redubur2}
U_{1,Y}+U_{1,T}+V_{1,Y}=0.
\end{equation}
Next, we substitute Eq. \eqref{redusol} into Eq. \eqref{ft} and vanishing different powers of $\tanh(\frac{x+F}{\sqrt{2b}})$ we get
\begin{equation}
(U_1-aV_1)F_Y+2baF_{YT}-(1+aV_1)F_T+b(a+1)F_{YY}+abF_{TT}+aV_1=0.
\end{equation}
Finally, we get three symmetry reduction equations by substituting Eq. \eqref{redusol} into Eq. \eqref{u1bur1} and vanishing different powers of $\tanh(\frac{x+F}{\sqrt{2b}})$, two of them are trivial leaving only the third one
\begin{multline}
(aV_1-U_1)U_{1,Y}+(aV_1-U_1)F_Y^2+[(1+aV_1)F_{T}-baF_{TT}-4baF_{YT}-3b(a+1)F_{YY}\\-aV_1]F_Y
+(1+aV_1)U_{1,T}-2ba(F_{YY}+F_{YT})F_T+b[2aF_{YT}-(a+1)U_{1,YY}\\-aU_{1,TT}
+2aF_{YY}-2aU_{1,YT}]=0
\end{multline}

The entrance of the tanh part in Eqs. \eqref{redusol} indicate the intrusion of an additional soliton to the Burgers wave. In other words, the group invariant solution
\eqref{redusol} is an interaction solution of one soliton and one general Burgers wave.

\section{B\"{a}cklund Transformations and Interaction Solutions of the Burgers Equation}
 The tanh function is the simplest automorphic function which possesses the property that its derivatives can be explicitly expressed by itself. Enlightened by the form of reduction solutions of Burgers equation in \eqref{redusol}, we now use the general tanh function expansion method to obtain interaction solutions between solitons and Burgers waves.

 The general truncated tanh expansion for (2+1)-dimensional Burgers system \eqref{bur} has the form
 \begin{subequations}\label{painex}
\begin{equation}\label{painex1}
u =w_1\tanh(\lambda t+\phi)+w_0
\end{equation}
\begin{equation}\label{painex2}
v =p_1\tanh(\lambda t+\phi)+p_0
\end{equation}
\end{subequations}
where $w_1,\, w_0,\, p_1,\, p_0$ and $\phi$ are arbitrary functions of $x,\,y$ and $t$. The $\lambda t$ in Eq. \eqref{painex} have been added for later convenience.

Now, substituting the tanh expansion \eqref{painex} into the Burgers system \eqref{bur} and vanishing different powers of $\tanh(\phi)$ in both expansion equations we can prove the following B\"{a}cklund transformation theorem.

\noindent\emph{ \textbf{Theorem 3.}} If $f_1$ and $f_2$ is the solution of the Burgers system \eqref{bur}, then so is $u$ and $v$ given by
\begin{subequations}\label{backlund}
\begin{equation}\label{backlund1}
u = 2b(\tanh(\phi+\lambda t)-1)\phi_y+f_1
\end{equation}
\begin{equation}\label{backlund2}
v = 2b(\tanh(\phi+\lambda t)-1)\phi_x+f_2
\end{equation}
\end{subequations}
with
\begin{equation}\label{phitlambda}
\phi_t=b\phi_{yy}-2b\phi_y^2+f_1\phi_y+ab\phi_{xx}-2ab\phi_x^2+af_2\phi_x-\lambda
\end{equation}

\emph{Proof}.
Substituting Eq. \eqref{painex} into Burgers system \eqref{bur} with $\lambda=0$ temporarily, we get
\begin{multline}\label{tanhbur1}
-w_1(2b\phi_y^2+2ab\phi_x^2-w_1\phi_y-ap_1\phi_x)\tanh^3(\phi)+(2bw_{1,y}\phi_y
+w_0w_1\phi_y+abw_1\phi_{xx}\\-w_1w_{1,y}+2abw_{1,x}\phi_x+ap_0w_1\phi_x-ap_1w_{1,x}
+bw_1\phi_{yy}-w_1\phi_t)\tanh^2(\phi)\\+(-abw_{1,xx}+2bw_1\phi_y^2-w_1w_{0,y}-bw_{1,yy}-w_0w_{1,y}
-ap_1w_1\phi_x-ap_1w_{0,x}-ap_0w_{1,x}-w_1^2\phi_y\\+2abw_1\phi_x^2+w_{1,t})\tanh(\phi)-2abw_{1,x}
\phi_x-abw_1\phi_{xx}+w_1\phi_t-2bw_{1,y}\phi_y+w_{0,t}-abw_{0,xx}\\-w_0w_{0,y}-ap_0w_1\phi_x
-bw_{0,yy}-w_0w_1\phi_y-ap_0w_{0,x}-bw_1\phi_{yy}=0
\end{multline}
and
\begin{multline}\label{tanhbur2}
(-w_1\phi_x+p_1\phi_y)\tanh^2(\phi)+(w_{1,x}-p_{1,y})\tanh(\phi)+w_1\phi_x+w_{0,x}-p_1\phi_y
-p_{0,y}=0.
\end{multline}
Vanishing the coefficient of $\tanh^3(\phi)$ in Eq. \eqref{tanhbur1} and $\tanh^2(\phi)$ in Eq. \eqref{tanhbur2} we get the solutions
\begin{equation}\label{w1p1}
w_1 = 2b\phi_y, p_1 = 2b\phi_x
\end{equation}
The coefficients of $\tanh^2(\phi)$ and  $\tanh(\phi)$ in Eq. \eqref{tanhbur1} are equivalent after using Eq. \eqref{w1p1}, which reads
\begin{equation}\label{phit}
\phi_t =b\phi_{yy}+w_0\phi_y+ab\phi_{xx}+ap_0\phi_x
\end{equation}
After using Eq. \eqref{w1p1}, Eq. \eqref{tanhbur2} is degenerated into the Burgers Eq. \eqref{bur2}
\begin{equation}
w_{0,x}-p_{0,y}=0
\end{equation}

Vanishing the coefficient of $\tanh^0(\phi)$ in Eq. \eqref{tanhbur1}, we get
\begin{equation}\label{fbur1}
f_{1,t}-f_1f_{1,y}-af_2f_{1,x}-bf_{1,yy}-abf_{1,xx}=0
\end{equation}
with
\begin{equation}\label{f1w0}
f_1 = w_0+2b\phi_y
\end{equation}
and
\begin{equation}\label{f2p0}
f_2 = p_0+2b\phi_x
\end{equation}
It can be easily verified that $f_1$ and $f_2$ also satisfy
\begin{equation}\label{fbur2}
f_{1,x}-f_{2,y}=0
\end{equation}
It is obvious that Eqs. \eqref{fbur1} and \eqref{fbur2} are just the (2+1)-dimensional Burgers
system. Substituting $w_0$ and $p_0$ by solving Eqs. \eqref{f1w0} and \eqref{f2p0} into Eq. \eqref{phit} meanwhile replacing $\phi$ by $\lambda t+\phi$ we get the Eq. \eqref{phitlambda}.

Now substituting $w_0,\, p_0,\, w_1$ and $p_1$ into Eq. \eqref{painex} by using Eqs. \eqref{w1p1},  \eqref{f1w0} and \eqref{f2p0} we finally get the B\"{a}cklund transformation \eqref{backlund}, then the theorem is proved.

If the seed functions are taken  as constants, i.e. $f_1=c_1$ and $f_2=c_2$, then Eq. \eqref{phitlambda} becomes a potential Burgers equation with source $\lambda$. Even for the seed solutions taken as constants, we can still obtain rich nontrivial solutions from the B\"{a}cklund transformation \eqref{backlund}. For instance, the straight line solution $\phi=k_1x+k_2y+\omega t$
of Eq. \eqref{backlund} leads to the single kink soliton solution.
\begin{subequations}
\begin{equation}\label{constsol1}
u = -2bk_2\{\tanh[k_2(2bk_2-c_1)t+ak_1(2bk_1-c_2)t-k_1x-k_2y]+1\}+c_1
\end{equation}
\begin{equation}
v = -2bk_1\{\tanh[k_2(2bk_2-c_1)t+ak_1(2bk_1-c_2)t-k_1x-k_2y]+1\}+c_2
\end{equation}
\end{subequations}
The traveling wave solution of Eq. \eqref{phitlambda} with $f_1=c_1$ and $f_2=c_2$ leads to a kink soliton solution
\begin{align}\label{phisol}
\phi =\nonumber-\frac{(K-k_2c_1-ak_1c_2+\omega)(k_1x+k_2y+\omega t-e_0)}{4b(k_2^2+ak_1^2)}\\-\frac{1}{2}\ln[1+\exp\big(\frac{-K(k_1x+k_2y+\omega t-e_1)}{b(k_2^2+ak_1^2)}\big)],
\end{align}
where $k_1,\,k_2,\,e_0,\,e_1$ and $\omega$ are arbitrary constants and K is related to other coantants by
\begin{equation}
K = \sqrt{k_2^2c_1^2-2k_2c_1\omega+2ak_1c_1k_2c_2+\omega^2-2ac_2k_1\omega+a^2c_2^2k_1^2-8k_2^2\lambda b-8\lambda abk_1^2}
\end{equation}
Substituting the kink soliton solution \eqref{phisol} into Eq. \eqref{backlund} with $f_1=c_1$ and $f_2=c_2$  we obtain the two soliton fusion solution for $u$ and $v$.

 \begin{figure}[htbp]
 \begin{center}
 \includegraphics[width=6cm,angle=-90]{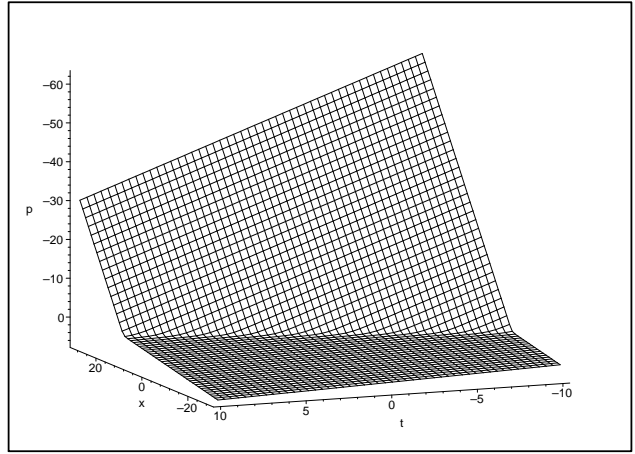}
 \caption{The kink soliton solution \eqref{phisol}.}
 \label{phikink}
 \end{center}
 \end{figure}

\begin{figure}[htbp]
 \begin{center}
 \includegraphics[width=6cm,angle=-90]{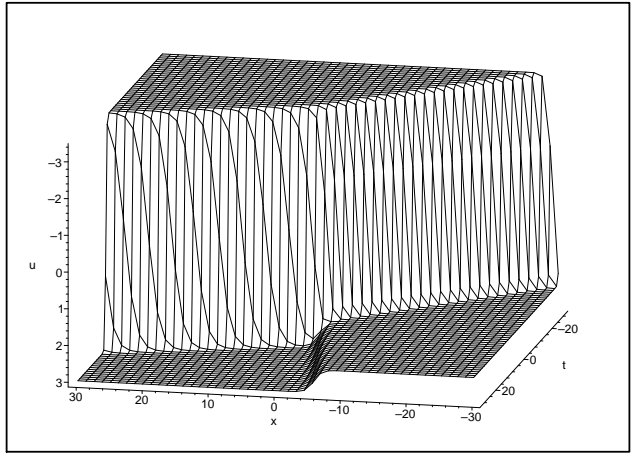}
 \caption{The two kink solitons fusion structure of $u$ in Eq. \eqref{backlund1} with Eq. \eqref{phisol}. }
 \label{ufusion}
 \end{center}
 \end{figure}

\begin{figure}[htbp]
 \begin{center}
 \includegraphics[width=6cm,angle=-90]{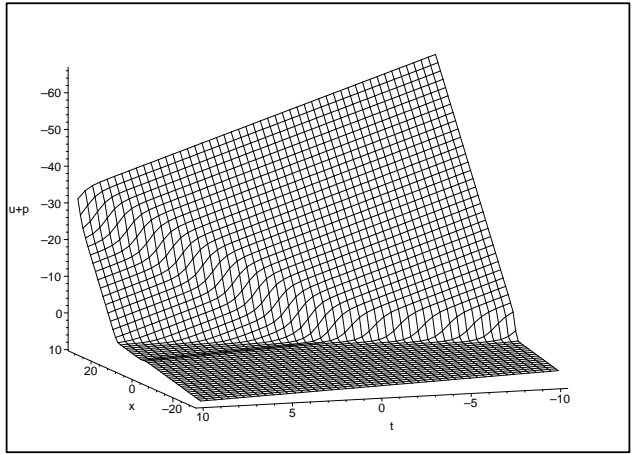}
 \caption{Evolution of three kink solitons fusion interaction.}
 \label{uphifusion}
 \end{center}
 \end{figure}
Now we draw the corresponding pictures respectively with the solutions \eqref{phisol}, \eqref{backlund1} with \eqref{phisol} and the interaction structure between them by plotting projectively at the $y=1$ plane while the parameters are fixed as
\begin{equation}\label{paramter1}
e_0=0,e_1=0,k_1=1,k_2=1,\omega=-1,a=1,b=1,c_1=3,c_2=3,\lambda=1.
\end{equation}

Figure \ref{phikink} reveals that solution \eqref{phisol} of Eq. \eqref{phitlambda} is a kink soliton. From  figure \ref{ufusion} which displays the soliton structure of Eq. \eqref{backlund1}, we see that two kink solitons fused to one at some time. Figure \ref{uphifusion} displays three soliton interaction structure between $u$ of \eqref{backlund1} with $f_1=c_1$ and $\phi$ of \eqref{phisol}. It is interesting to find from this picture that three kink solitons fused to one after interaction. Because the field $v$ of \eqref{backlund2} resemble $u$ in form, the interaction structure between the field $v$ with $\phi$ of \eqref{phisol} has the similar form as that showing in Fig. \ref{uphifusion}.

\section{Conclusion and discussion}
In summary, the (2+1)-dimensional Burgers equation is investigated by multiple ways to find interactions among different nonlinear excitations. The nonlocal residual symmetries which corresponds to the generators of truncated Painlev\'{e} expansions has been localized in an new prolonged system. The new BT 2 and symmetry reduction solutions thus has been obtained by using classical Lie point symmetry approach. The reduction solutions reveals that new soliton has been introduced to interact with background waves.  Enlightened by the form of reduction solutions of Burgers equation, we further studied this system by a new generalized tanh expansion method. With this method we give a new BT theorem 3 for Burgers system. It is interesting that from BT 3 we can get rich soliton structure even for the constant seed. Especially, two kink solitons fusion structure and evolution of three kink solitons fusing to one has been displayed by plotting projectively. This fact implies that starting from any seed solution of an integrable model, the BT 3 will introduce an additional soliton to the original seed.

\begin{acknowledgments}
The authors are in debt to Prof. S.Y. Lou for his valuable comments and suggestions. This work was supported by the National Natural Science Foundation of China under Grant No. 10875078, 11305106 and the Natural Science Foundation of Zhejiang Province of China Grant No. Y7080455, LQ13A050001.
\end{acknowledgments}

\end{document}